\documentclass{article}
\usepackage{cite}
\usepackage{graphicx}
\usepackage{dcolumn}

\begin{document}

\date{}
\title{Comment on: ``Bound states for a Coulomb-type potential induced by the
interaction between a moving electric quadrupole moment and a magnetic
field''. Ann. Phys. 341 (2014) 86}
\author{Francisco M. Fern\'{a}ndez\thanks{%
fernande@quimica.unlp.edu.ar} \\
%EndAName
INIFTA, DQT, Sucursal 4, C.C 16, \\
1900 La Plata, Argentina}
\maketitle

\begin{abstract}
We analyze the results obtained from a model consisting of the interaction
between a moving electric quadrupole moment and a magnetic field. We argue
that there are no bound states contrary to what the author stated. It is
shown that the proposed method is unsuitable for obtaining all the bound
states and that the existence of allowed cyclotron frequencies is an
artifact of the approach.
\end{abstract}

In a paper published in this journal Bakke\cite{B14} discussed a model that
consists of the interaction of a moving electric quadrupole moment and a
magnetic field and also added a two-dimensional harmonic-oscillator
potential. From the time-dependent Schr\"{o}dinger equation the author
derived an eigenvalue equation for the radial part of the wavefunction. He
solved the latter by means of the Frobenius method and an arbitrary
truncation of the power series suggested by the three-term recurrence
relation for the expansion coefficients. From the analysis of the particular
bound states thus obtained he drew the remarkable conclusion that not all
the oscillator frequencies are allowed. The purpose of this Comment is the
analysis of the approach used by Bakke\cite{B14} to obtain the bound states
of the Schr\"{o}dinger equation just mentioned and the validity of the
results thus derived.

After taking into account the interaction of a moving electric quadrupole
moment and a magnetic field the author derived the Schr\"{o}dinger equation
\begin{equation}
i\frac{\partial \psi }{\partial t}=-\frac{1}{2m}\left( \frac{\partial ^{2}}{%
\partial \rho ^{2}}+\frac{1}{\rho }\frac{\partial }{\partial \rho }+\frac{1}{%
\rho ^{2}}\frac{\partial ^{2}}{\partial \varphi ^{2}}+\frac{\partial ^{2}}{%
\partial z^{2}}\right) \psi -i\frac{Q\lambda _{m}}{2m\rho }\frac{\partial
\psi }{\partial \varphi }+\frac{Q^{2}\lambda _{m}^{2}}{8m}\psi ,
\label{eq:Schro_fime_dep}
\end{equation}
in cylindrical coordinates ($0\leq \rho <\infty $, $0\leq \varphi \leq 2\pi $%
, $-\infty <z<\infty $) and units such that $\hbar =c=1$ (we have
recently criticized this kind of nonrigorous choice of suitable
units\cite{F20}). The
meaning of the parameters in this equation is given in the author's paper%
\cite{B14}.

Since the Hamiltonian operator is time-independent and commutes with $\hat{p}%
_{z}$ and $\hat{L}_{z}$ the author looked for a particular solution of the
form
\begin{equation}
\psi (t,\rho ,\varphi ,z)=e^{-i\mathcal{E}t}e^{il\varphi }e^{ikz}R(\rho ),
\label{eq:psi}
\end{equation}
where $l=0,\pm 1,\pm 2,\ldots $ and $-\infty <k<\infty $. The function $%
R(\rho )$ satisfies the differential equation
\begin{equation}
R^{\prime \prime }+\frac{1}{\rho }R^{\prime }-\frac{l^{2}}{\rho ^{2}}R-\frac{%
\delta }{\rho }R+\zeta ^{2}R=0,  \label{eq:dif_eq_R_1}
\end{equation}
where
\begin{eqnarray}
\zeta ^{2} &=&2m\mathcal{E}-k^{2}-\frac{Q^{2}\lambda _{m}^{2}}{4},  \nonumber
\\
\delta  &=&Q\lambda _{m}l.  \label{eq:zeta,delta}
\end{eqnarray}

After analyzing the behaviour of $R(\rho )$ at infinity the author argued
that ``Therefore, we can find either scattering states $R\cong e^{i\zeta
\rho }$ or bound states $R\cong e^{-\tau \rho }$ if we consider $\zeta
^{2}=-\tau ^{2}$. Note that the fourth term on the left-hand side of Eq. (8)
plays the role of a Coulomb-like potential. This term stems from the
interaction between the magnetic field (5) and the electric quadrupole
moment defined in Eq (4). Our intention is to obtain bound state solutions,
and then, the term proportional to $\delta $ behaves like an attractive
potential by considering the negative values of $\delta $, that is, $\delta
=-|\delta |$. This occurs by considering either $\lambda _{m}>0$ and $l<0$
or $\lambda _{m}<0$ and $l>0$ (we consider $Q$ being always a positive
number). Observe that these conditions forbid the quantum number $l$ to have
the value $l=0$, that is, for $l=0$ there are no bound states solutions.''

This analysis is obviously wrong because the Schr\"{o}dinger equation
discussed above does not have bound-states for any value of $\delta $ since
the motion is unbounded along the $z$ direction (the Hamiltonian operator
commutes with $\hat{p}_{z}$). For this reason one introduces the term $%
e^{ikz}$, and the energy, which depends on $k^{2}/(2m)$, takes all the
values $\mathcal{E}\geq \zeta ^{2}/(2m)+Q^{2}\lambda _{m}^{2}/(8m)$. To be
more specific, we have bound states only if
\begin{equation}
\int \int \int \left| \psi (t,\rho ,\varphi ,z)\right| ^{2}\rho \,d\rho
\,d\varphi \,dz<\infty ,  \label{eq:bound-state_def}
\end{equation}
as shown in any textbook on quantum mechanics\cite{CDL77}. In all the
examples discussed here the improper integral over $z$ is obviously
divergent.

In the second model the author added the potential $V(\rho )=\frac{1}{2}%
m\omega ^{2}\rho ^{2}$ ($\rho ^{2}=x^{2}+y^{2}$) that is clearly unable to
bound the motion along the $z$ axis. Consequently, this model does not have
bound states and the spectrum is continuous as in the preceding one. By
means of the change of variables $\xi =\sqrt{m\omega }\rho $ one obtains the
radial eigenvalue equation
\begin{equation}
R^{\prime \prime }+\frac{1}{\xi }R^{\prime }-\frac{l^{2}}{\xi ^{2}}R-\frac{%
\alpha }{\xi }R-\xi ^{2}R+\frac{\zeta ^{2}}{m\omega }R=0,
\label{eq:dif_eq_R_2}
\end{equation}
where $\alpha =\frac{\delta }{\sqrt{m\omega }}$. There is a misprint in the
sign of the term $\alpha /\xi $ in the author's equation (18) that also
appears in his equation (21).

On writing the solution $R(\xi )$ as
\begin{equation}
R(\xi )=\xi ^{|l|}e^{-\frac{\xi ^{2}}{2}}\sum_{j=0}^{\infty }a_{j}\xi ^{j},
\label{eq:R_series}
\end{equation}
one obtains the three-term recurrence relation
\begin{eqnarray}
a_{j+2} &=&\frac{\alpha }{(j+2)(j+1+\theta )}a_{j+1}-\frac{g-2j}{%
(j+2)(j+1+\theta )}a_{j},  \nonumber \\
j &=&-1,0,1,2\ldots ,\;a_{-1}=0,\;a_{0}=1,  \label{eq:rec_rel}
\end{eqnarray}
where $\theta =2|l|+1$ and $g=\frac{\zeta ^{2}}{m\omega }-2-2|l|$. The
author did not carry the misprint in the sign of $\alpha $ to this equation.

The author argued as follows: ``Bound states solutions correspond to finite
solutions; therefore, we can obtain bound states solutions by imposing that
the power series expansion (23) or the Heun biconfluent series becomes a
polynomial of degree $n$. Through the expression (24), we can see that the
power series expansion (23) becomes a polynomial of degree $n$ if we impose
the conditions:
\begin{equation}
g=2n\ \mathrm{ and }\ a_{n+1}=0,  \label{eq:truncation_cond}
\end{equation}
where $n=1,2,\ldots $.''

It clearly follows from these two conditions that $a_{j}=0$ for all $j>n$;
however, the author's statement is a gross conceptual error because a bound
state simply requires that $R(\xi )$ is square integrable:
\begin{equation}
\int_{0}^{\infty }\left| R(\xi )\right| ^{2}\xi d\xi <\infty .
\label{eq:bound-state_def_xi}
\end{equation}
Before proceeding with the discussion of Bakke's paper we assume that the
motion of the particle is restricted to the $x-y$ plane so that we can truly
speak of bound states, otherwise such states do not exist as discussed above.

For example, when $n=1$ then $g=2$ and we obtain a simple analytical
expression for $\mathcal{E}_{1,l}$. The second condition $a_{2}=0$ yields $%
\alpha =\alpha _{1,l}=\pm \sqrt{4|l|+2}$ and the cyclotron frequency $\omega
_{1,l}=\delta ^{2}/(m\alpha _{1,l}^{2})$. From the general case (\ref
{eq:truncation_cond}) the author derived an analytic expression for $%
\mathcal{E}_{n,l}$ corresponding to $\omega _{n,l}=\delta ^{2}/(m\alpha
_{n,l}^{2})$. Accordingly, he argued as follows: ``Hence, we have seen in
Eq. (29) that the effects of the Coulomb-like potential induced by the
interaction between the magnetic field and the electric quadrupole moment on
the spectrum of energy of the harmonic oscillator correspond to a change of
the energy levels, where the ground state is defined by the quantum number $%
n=1$ and the angular frequency depends on the quantum numbers $\{n,l\}$.
This dependence of the cyclotron frequency on the quantum numbers $\{n,l\}$
means that not all values of the cyclotron frequency are allowed, but a
discrete set of values for the cyclotron frequency.''

Both conclusions are wrong as we shall see in what follows. In the first
place, the truncation conditions (\ref{eq:truncation_cond}) only yield some
rather rare bound states, based on polynomial functions, that occur for some
arbitrary particular values of $\omega $. Notice that we can also solve $%
\alpha =\alpha _{n,l}$ for $Q$ (or for $\lambda _{m}$) for arbitrary values
of $\omega $, in which case case either $Q$ or $\lambda _{m}$ will be
quantized instead of $\omega $. More precisely, quantization of the model
parameters is an artifact of looking for exact solutions of any
quasi-solvable (or conditionally solvable) model\cite{D88, BCD17} (and
references therein).

In order to prove the conclusion about the ground state wrong, we solve the
eigenvalue equation (\ref{eq:dif_eq_R_2}) for $W=\frac{\zeta ^{2}}{m\omega }$
by means of the reliable Rayleigh-Ritz variational method with the basis set
of (unnormalized) functions $\left\{ u_{j}(\xi )=\xi ^{|l|+j}e^{-\frac{\xi
^{2}}{2}},\;j=0,1,\ldots \right\} $. This approach is well known to yield
increasingly accurate upper bounds to all the eigenvalues of the Schr\"{o}%
dinger equation\cite{P68} (and references therein). We tested the accuracy
of these results by means of the powerful Riccati-Pad\'{e} method\cite
{FMT89a}. As a suitable particular case we choose $l=1$ and $\alpha =\alpha
_{1,1}=\sqrt{6}$ so that the supposed ground-state eigenvalue obtained by
Bakke is $W=6$. For the first three eigenvalues both methods yield $%
W_{0,1}=1.600357154$, $W_{1,1}=6$, $W_{2,1}=10.21072810$. We appreciate that
the author did not obtain the ground-state energy but the first-excited one
and present numerical calculations confirm the fact that the quantization of
the model parameters is an artifact of the truncation method used to solve
the eigenvalue equation.

Since the transformation $(\alpha ,\xi )\rightarrow (-\alpha ,-\xi )$ leaves
the eigenvalue equation (\ref{eq:dif_eq_R_2}) unchanged, we conclude that $%
W(-\alpha )=W(\alpha )$ (or $W_{n,l}=W_{n,-l}$). The two calculations
mentioned above with $\alpha =\alpha _{1,-1}=-\sqrt{6}$ confirm that $%
W_{n,-1}=W_{n,1}$.

In two earlier papers Bakke and Belish\cite{BB12} and Bakke and Moraes\cite
{BM12} discussed the quantization of $\omega $ and $\omega $ or $k$,
respectively, in quantum-mechanical models that lead to exactly the same
differential equation and resorting to exactly the same strategy based on
truncation of a power series that leads to a three-term recurrence relation.

Summarizing: Contrary to what Bakke presumed, none of the two models
discussed supports bound states because the motion of the particle
(quadrupole moment) along the $z$ axis is unbounded. Bound states occur only
if we restrict the motion in the $x-y$ plane. In the second model (under the
latter assumptions) the approach followed by the author only yields some
extremely particular states for also extremely particular model parameters.
Anybody familiar with the Schr\"{o}dinger equation realizes that for any
real value of $\alpha $ in the equation (\ref{eq:dif_eq_R_2}) one obtains an
infinite set of eigenvalues $W_{n,l}(\alpha )$. The truncation method
proposed by the author is unsuitable for obtaining the spectrum except for
some particular values of $\alpha $ and in each of these cases it provides
just one eigenvalue. For example, for $\alpha =\pm \sqrt{6}$ it only yields $%
W_{1,\pm 1}=6$ and certainly misses all the other eigenvalues (we have
already shown two more above). In addition to all this, the predicted
existence of allowed cyclotron frequencies is merely an artifact of the
truncation method proposed by Bakke.

\end{document}